\documentclass[sigconf,natbib=true]{acmart}
\setcopyright{none}
\renewcommand\footnotetextcopyrightpermission[1]{} % removes footnote with conference information in first column
\usepackage{bbm}
\usepackage{cleveref}
\usepackage{multirow}
\usepackage{multicol}
\usepackage{graphicx}

\newcommand{\squishlist}{
 \begin{list}{$\bullet$}
  { \setlength{\itemsep}{0pt}
     \setlength{\parsep}{3pt}
     \setlength{\topsep}{3pt}
     \setlength{\partopsep}{0pt}
     \setlength{\leftmargin}{1.5em}
     \setlength{\labelwidth}{1em}
     \setlength{\labelsep}{0.5em}}}

\newcommand{\squishlisttwo}{
 \begin{list}{$\bullet$}
  { \setlength{\itemsep}{0pt}
     \setlength{\parsep}{0pt}
    \setlength{\topsep}{0pt}
    \setlength{\partopsep}{0pt}
\setlength{\leftmargin}{2em}
\setlength{\labelwidth}{1.5em}
\setlength{\labelsep}{0.5em} } }

\newcommand{\squishend}{
\end{list}  }

\AtBeginDocument{%
  \providecommand\BibTeX{{%
    \normalfont B\kern-0.5em{\scshape i\kern-0.25em b}\kern-0.8em\TeX}}}

\acmConference[RecSys 22]{September 18--23,
  2022}{Seattle, Washington}
    
\begin{document}

\author{Allen Lin}
\affiliation{%
  \institution{Texas A\&M University}
  \city{College Station, Texas}
  \country{USA}
}

\author{Ziwei Zhu}
\affiliation{%
  \institution{George Mason University}
  \city{Fairfax, Virginia}
  \country{USA}
}

\author{Jianling Wang}
\affiliation{%
  \institution{Texas A\&M University}
  \city{College Station, Texas}
  \country{USA}
}

\author{James Caverlee}
\affiliation{%
  \institution{Texas A\&M University}
  \city{College Station, Texas}
  \country{USA}
}

%\author{Shuo Lin, Jianling Wang, Ziwei Zhu, James Caverlee}
%\affiliation{%
%  \institution{Texas A\&M University}
%  \city{College Station, Texas}
%  \country{USA}
%}

%%
%% The "title" command has an optional parameter,
%% allowing the author to define a "short title" to be used in page headers.
\title{Towards Fair Conversational Recommender Systems}

%%
%% The abstract is a short summary of the work to be presented in the
%% article.
\begin{abstract}
Conversational recommender systems have demonstrated great success. They can accurately capture a user’s current detailed preference -- through a multi-round interaction cycle -- to effectively guide users to a more personalized recommendation. Alas, conversational recommender systems can be plagued by the adverse effects of bias, much like traditional recommenders. In this work, we argue for increased attention on the presence of and methods for counteracting bias in these emerging systems. As a starting point, we propose three fundamental questions that should be deeply examined to enable fairness in conversational recommender systems.
\end{abstract}

\maketitle

\section{Introduction}
%Ranging from E-commerce, to multimedia services, and to E-education platforms, recommender systems alleviate information overload by 

%Recommender systems have become an indispensable tool in everyone's daily life, connecting users to their items of interest. While traditional recommender systems have been widely adopted, they have two inherent disadvantages. First, the learning process of traditional recommender systems is often  based on historical interaction data which can be sparse, noisy, and not reflective of the current interests of the user. Therefore, it is difficult for traditional recommender systems to \textit{precisely capture the current preference of a user}. Second, the recommendations made by these traditional systems \textit{can be difficult to justify} since the user's preference towards an item is typically based on learned representations that are uninterpretable to humans. 

\par Recently, a new form of interactive recommendation system -- the  conversational recommender system (CRS) --  has shown great success in enhancing personalization through multi-turn interactions (i.e., dialogue) \cite{DBLP:journals/corr/abs-2002-09102, DBLP:journals/corr/abs-1806-03277, DBLP:journals/corr/abs-2007-00194, xu2021adapting, deng2021unified, hu2022learning}. Compared to traditional recommender systems, a CRS can better capture the current preference of a user through direct elicitation and provide natural justifications for its recommendations. However, along with the promising functionality and wide range of potential applications like movies  \cite{li2018towards, zhou2020towards}, music \cite{xu2021adapting, hu2022learning}, and e-commerce \cite{deng2021unified}, the move to conversation also opens the door for new challenges.

%As a result, CRSs have been applied to movies  \cite{li2018towards, zhou2020towards}, music \cite{xu2021adapting, hu2022learning}, and e-commerce \cite{deng2021unified}.
%can In this way, the system overcomes the above mentioned two difficulties by eliciting users' current and detailed preferences, leading to justifiable and highly personalized recommendations. Given its promising functionality, CRSs 

%Despite there has been many works on CRSs, the fairness aspect remains significantly understudied.   

\par For example, a few recent works have begun to study some of these challenges, including issues like \textit{multi-turn conversational strategies and exploration}~\cite{DBLP:journals/corr/abs-1806-03277, 10.1145/3269206.3271776, 10.1145/3437963.3441791} and \textit{exploitation trade-offs}~\cite{zhang2020conversational, yu2019visual, DBLP:journals/corr/abs-2005-12979, christakopoulou2016towards}. We are encouraged by this attention, and argue here that conversational recommender systems demand new inquiries to better understand the fairness aspect as these systems become more widely adopted. In particular, we argue that fairness in CRSs should examine the three following questions:
\squishlist
\item \textbf{To what degree does bias exist in CRSs? And how should we quantify and mitigate it?} Since a CRS directly elicits user preference, one could argue that the recommendations made are completely unbiased since they are all strictly based upon the user's specified preferences during the conversation. Yet our preliminary work in this space \cite{lin22cikm} shows how popularity bias can still arise; what other forms of bias are there as well?% forms of bias can still arise. What other forms of bias exist?

%Since CRSs also take into consideration the past user-item interactions when making recommendations, it would be reasonable to assume that CRSs are affected by the adverse effect of bias similar to traditional systems. 

\item \textbf{Could the preference elicitation process itself be biased?} User preference elicitation lies in the core of CRSs. Through multi-turn interactions, CRSs engage with users (for example, through a question-and-answer paradigm) to precisely elicit the current and detailed preferences of the user. Given its importance, this elicitation process itself must be carefully studied. For example, does a CRS bias towards certain attributes of either items or users in how it elicits user preference?%unbiased and highly personalized instead of prompting only on a subset of the entire attribute space to the user.

\item \textbf{How should we evaluate a CRS in an unbiased way?} Given the added user preference elicitation process, the evaluation of a CRS poses new challenges. Many previous works \cite{DBLP:journals/corr/abs-2002-09102, DBLP:journals/corr/abs-1806-03277, DBLP:journals/corr/abs-2007-00194, xu2021adapting, deng2021unified, hu2022learning} have adopted metrics like Success Rate (SR@t) to evaluate the quality of a CRS. However, we argue that this metric may not necessarily result in a fair evaluation of the system, and so further study is needed.   
\squishend

\section{Preliminaries}
\label{sec:prelim}
Before proceeding, we provide a brief introduction for the workflow of a CRS. We follow the commonly adopted System Ask - User Responds (SAUR) setting \cite{10.1145/3269206.3271776} to investigate  fairness in a multi-round CRS. Formally let \begin{math} U \end{math} denote the user set, \begin{math} V \end{math} denote the itemset, and \begin{math} A = a_1,a_2,...,a_m \end{math} denote a set of m domain-specific attributes used to systematically characterize all items in \begin{math} V \end{math}. At each turn, a CRS calls upon its conversation component  to decide whether to ask the user's preference on a specific attribute or make a recommendation. If this policy agent decides not enough preference evidence has been collected, it will pick one attribute \begin{math} a \end{math} from the set of unasked attributes to prompt the user. When prompted with the question, the user is assumed to provide her preference \begin{math} p \end{math} on the asked attribute \begin{math} a \end{math}. Upon receiving the user's preference, the policy agent updates its current belief state \begin{math} s \end{math} by adding the (attribute, preference) pair to it. If the policy agent decides enough information has been collected after t turns of interaction, the CRS then calls upon its recommender component to make a list of recommendations. Unlike a  static recommender which ranks all items in the entire itemset, the recommender component of a CRS only ranks items in the candidate itemset \begin{math} V_{cand} \end{math} -- the itemset that contains only items with attributes perfectly matching all (attribute, preference) pairs in the current belief state. After ranking all items in \begin{math} V_{cand} \end{math}, the CRS recommends the top K items to display to the user. If the user accepts the recommendation, then the system quits. If the user rejects all the recommended items, the system repeats the cycle by calling upon its policy agent to decide the next action to take. This process continues until the user quits due to impatience or a predefined maximum number of turns has been reached.    

\begin{figure*}
  \includegraphics[width=1.01\textwidth]{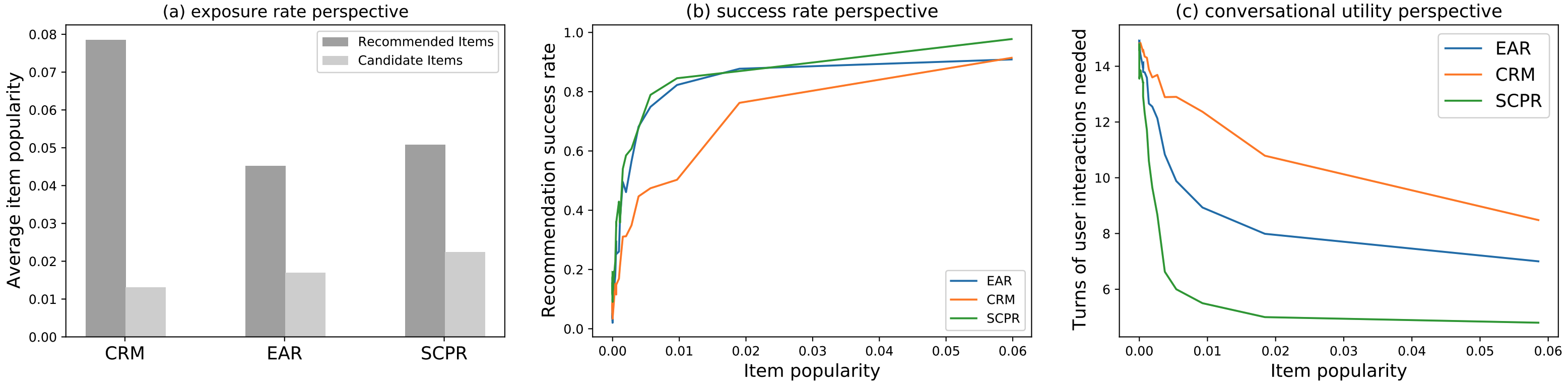}
  \caption{Evidence of different perspectives of popularity bias in CRSs}
  \Description{Evidence of Popularity bias in CRSs.}
  \label{fig:show bias}
\end{figure*}

\section{Evidence of Bias in CRS}
\label{sec:evidence}
In our preliminary work \cite{lin22cikm}, we have begun to study \textit{popularity bias} in the context of CRSs. Specifically we conducted data-driven analysis on the Lastfm dataset for music artist recommendation with three state-of-the-art CRSs: SCPR~\cite{DBLP:journals/corr/abs-2007-00194}, EAR~\cite{DBLP:journals/corr/abs-2002-09102}, and CRM~\cite{DBLP:journals/corr/abs-1806-03277}, to illustrate the existence of popularity bias in a multi-round CRS from three different perspectives:

\smallskip
\noindent\textbf{\textit{Exposure Rate} Perspective.} A CRS starts by collecting a set of user preferences, from which the system forms a candidate item set and then recommends the top-k items. As shown in Figure~\ref{fig:show bias}(a), the average popularity of items that get recommended by an example system like CRM is almost 8 times higher than the average popularity of items in the candidate set. A similar pattern is observed in EAR and SCPR. However, in a completely unbiased CRS, since both the candidate set and the recommendation set contain items perfectly matching all preferences specified by the user (i.e., equally qualified items), the overall average popularity for both sets should not deviate excessively.  

% containing all items matching the collected user preferences so far. Once the system determines enough information has been collected, it selects the top N items in the candidate set to generate the recommendation set.

\smallskip
\noindent\textbf{\textit{Success Rate} Perspective.} In a completely unbiased system, an item's popularity should be irrelevant to its recommendation success rate. Any correlation, either positive or negative, is an indication of popularity bias in the CRS. As we can observe from Figure~\ref{fig:show bias}(b), all three models exhibit strong positive correlations between an item's popularity and its recommendation success rate. For example, in EAR, items with popularity greater than 0.01 -- the 75th percentile of the sorted popularity of all items -- have over 75\% recommendation success rate. That is, items with high popularity have much higher chances of being successfully recommended.

\smallskip
\noindent\textbf{\textit{Conversational Utility} Perspective.} Unlike traditional recommenders, a CRS operates under a multi-round interaction setting, in which a user might leave due to impatience. Intuitively, items requiring fewer turns of user interactions  to get successfully recommended should be considered more advantageous than those requiring more. Thus in an unbiased CRS, an item's required turns of user interactions should be completely independent of its popularity. However, as shown in Figure~\ref{fig:show bias}(c), all three models exhibit strong inverse correlations between an item's popularity and its required turns of user interaction to get successfully recommended.

\smallskip
The evidence here is around the issue of popularity bias, but what other types of bias arise? And how can we begin to mitigate these types of bias? Are traditional de-biasing methods appropriate in a conversational setting? And if not, what steps should we take? In the following we highlight two special kinds of bias that arise in conversational recommenders.

%While we have proposed ways to quantify and mitigate popularity bias in CRSs in our initial work \cite{x}, further studies are yet to be conducted to investigate the existence of other types of bias in CRSs. One such example is the user preference assumption bias. 

\section{User Preference Assumption Bias}
A CRS starts by prompting users with questions to collect their current and detailed preferences. This process is called user preference elicitation. In each turn, the system can choose to ask the user's preference on one (or more) of the domain-specific attributes that systematically characterize the itemset. The recorded user responses are then used to help the system to ideally make more personalized recommendations. \textit{User Preference Assumption Bias} refers to the phenomenon of the user preference elicitation process being dominated by the system only asking the user's preferences on a small subset of the entire domain-specific attributes. Take movie recommendation as an example, since the attribute \textit{length} can help to quickly narrow down the candidate item space, the CRS would assume that every user has a strong preference to towards specifying the length of the movie. However, in reality, this might not always be true. We argue that the questions prompted to each user should be highly personalized and remain uninfluenced by the false preference assumptions that the system makes about the user.  

\section{Recommendation Evaluation Bias}
To evaluate a CRS, many previous works \cite{DBLP:journals/corr/abs-2002-09102, DBLP:journals/corr/abs-1806-03277, DBLP:journals/corr/abs-2007-00194, xu2021adapting, deng2021unified, hu2022learning} have adopted Success Rate (SR@t) as the evaluation metric. SR@t measures the ratio of successful conversations, i.e., if the system is able to recommend the ground truth item by turn t. While the intuition seems reasonable, this metric does not evaluate the system in an unbiased fashion. SR@t is defined as \begin{math} \frac{\# \textit{successfully recommended items @ t}}{\textit{total number of items}}\end{math}. Assume we only have 2 users: Alice is extremely active and has 90 ground truth items in the testing set while Bob only has 10. Now consider two scenarios: scenario a, the system successfully recommends all 90 items, within the allowed turns of interactions t, for Alice and 0 for Bob vs. scenario b, the system successfully recommends 81 items for Alice and 9 items for Bob. The SR@t calculated would be identical in both scenarios. However from the fairness perspective, scenario b provides much better results for both users than scenario a. 

\section{Conclusion}
In this position paper, we address the need for further studies in the fairness aspect of CRSs and propose three fundamental questions that we hope the community will continue to examine.

\bibliographystyle{ACM-Reference-Format}
\bibliography{ref}

\end{document}